\documentclass[prc,aps,twocolumn,floats,floatfix,showpacs,superscriptaddress,nofootinbib]{revtex4}
\usepackage{amsmath}
\usepackage{amsfonts}
\usepackage{graphicx}

\begin{document}

\title{Multi-neutron emission of Cd isotopes}

\author{A. P. Severyukhin}
\affiliation{Bogoliubov Laboratory of Theoretical Physics,
             Joint Institute for Nuclear Research,
             141980 Dubna, Moscow region, Russia}
\affiliation{Dubna State University, Universitetskaya street 19,
             141982 Dubna, Moscow region, Russia}
\author{N. N. Arsenyev}
\affiliation{Bogoliubov Laboratory of Theoretical Physics,
             Joint Institute for Nuclear Research,
             141980 Dubna, Moscow region, Russia}
\author{I. N. Borzov}
\affiliation{National Research Centre "Kurchatov Institute",
             123182 Moscow, Russia}
\affiliation{Bogoliubov Laboratory of Theoretical Physics,
             Joint Institute for Nuclear Research,
             141980 Dubna, Moscow region, Russia}
\author{E. O. Sushenok}
\affiliation{Bogoliubov Laboratory of Theoretical Physics,
             Joint Institute for Nuclear Research,
             141980 Dubna, Moscow region, Russia}
\affiliation{Dubna State University, Universitetskaya street 19,
             141982 Dubna, Moscow region, Russia}

\begin{abstract}
An influence of the phonon-phonon coupling (PPC) on the
$\beta$-decay half-lives and multi-neutron emission probabilities
is analysed within the microscopic model based on the
Skyrme interaction with tensor components included. The
finite-rank separable approximation is used in order to handle
large two-quasiparticle spaces. The even-even nuclei near the
r-process pathes at $N=82$ are studied. The characteristics of
ground states, $2^+$ excitations and $\beta$-decay strength of the
neutron-rich Cd isotopes are treated in detail. It is shown that a
strong redistribution of the Gamow-Teller strength due to the PPC
is mostly sensitive to the multi-neutron emission probability of
the Cd isotopes.
\end{abstract}


\date{\today}

\maketitle
%
%=======================================================
%
\section{Introduction}
%
%=====================================================================
%
The $\beta$-decay properties are very important for understanding
the nuclear structure evolution at extreme $N/Z$ ratios,
for analysis of the radioactive ion-beam experiments and
modeling of astrophysical $r$-process~\cite{LM03}. In the last
years a renewed attention has been attracted to the delayed
multi-neutron emission ($\beta x n$) with $x = 2, 3 ...$. The
$\beta 2 n$ emission has been predicted in early 60-s~\cite{G60}
and later observed in for the cases of $^{11}$Li~\cite{A79} and
$^{30-32}$Na~\cite{D80}.
 It has also been considered for heavier nuclei
in Ref.~\cite{L85} emphasizing a competition between the sequential
and resonant (''di-neutron'') emission. Observation of the
di-neutron emission in $^{16}$Be decay has been recently
claimed~\cite{S12} (see comment in~\cite{M12} and discussion in~\cite{B16}).
Now a days Bi isotopes in the mass region $N > 126$ are the heaviest nuclei
where the delayed neutron emission has been studied~\cite{C16}.

A study of $\beta x n$ processes facilitates developing a
self-consistent approach based on the energy density functional
(EDF). The process probability depends, first on specific energy
``landmarks'': the $\beta$-decay energy release $Q_{\beta}$ and
neutron emission thresholds $S_{xn}$. An adequate description of
these differential quantities poses constraints on the EDF at high
isospin-asymmetry regime. Second crucial ingredient is the
$\beta$-strength function: spectral distribution of the
$\beta$-decay matrix elements within the $\beta$-decay window
($Q_{\beta}$). Assuming an amplification of the intensity
distribution by the integral Fermi factor, the most important
contributions come from the allowed Gamow-Teller~(GT) and
high-energy first-forbidden $\beta$-decays. Importantly,
they should be treated employing the one
and the same self-consistent framework~\cite{ff03,pn05}.

Experimental studies using the multipole decomposition analysis of
the (n,p) and (p,n) reactions~\cite{w97,mda} found substantial GT
strength above the GT resonance peak.
This solves a longstanding problem
of the missing experimental GT strength.
Also it helps to overcome the
discrepancies between the theoretical predictions
using the one-phonon wave-function of
the quasiparticle random phase approximation (QRPA) and the
measurements. It has been found necessary to take into account the
coupling with more complex configurations in order to shift some
strength to higher transition energies
in order to comply with the experimental results
\cite{bh82,KS84,dosw87}. Using the Skyrme EDF and the RPA, such
attempts in the past~\cite{cnbb94,csnbs98} have allowed one to
understand the damping of charge-exchange resonances and their
particle decay. The damping of the GT mode has been investigated
using the Skyrme-RPA plus particle-vibration
coupling~\cite{ncbbm12}. The main difficulty is that
complexity of the calculations increases rapidly
with the size of the configurational space and one has to work
within limited spaces. The separable form of the residual
interaction is the practical advantage of the quasiparticle phonon
model (QPM)~\cite{solo} which allows one to perform the
calculations in large configurational
spaces~\cite{KS84,KS85,solo}. The finite rank separable
approximation (FRSA) for the QRPA with Skyrme
interactions~\cite{gsv98,svg04} has been invented  to describe
charge-exchange excitation modes~\cite{svg12,svbag14}.

In the present paper we concentrate on the delayed multi-neutron
emission in the region below the neutron-rich doubly-magic nucleus
$^{132}$Sn. The $\beta$-decay properties of $r$-process
``waiting-point nuclei'' $^{129}$Ag, $^{130}$Cd, and $^{131}$In
have attracted a lot of experimental efforts
recently~\cite{H01,D03,L15,D16,J16}. The theoretical analysis has
been done within the microscopic-macroscopic finite-range droplet
model~(FRDM+QRPA)~\cite{H01,M97}, the continuum QRPA approach with
the Fayans EDF (DF3+cQRPA)~\cite{ff03,pn05,pn06,B08}. Recently, the
proton-neutron relativistic QRPA~(pn-RQRPA)~\cite{Marketin16} and
the finite-amplitude method~(FAM)~\cite{Mustonen16} calculations
have appeared. In general, the microscopic
approaches~\cite{pn05,Marketin16,Mustonen16} described the
half-lives and total probabilities of the $\beta x n$ emission
better than the global approach~\cite{M03} commonly used for
astrophysical $r$-process modeling. Importantly, all the cited
papers have used the one-phonon approximation. This may be
not enough
for adequate reproduction of the fine structure of the GT
strength distribution near the neutron thresholds. Such a detailed
analysis is feasible in the large-scale shell-model~\cite{Z13} but
this approach is limited
by the number of available $np-nh$ configurations.

In most of the cases, the experimental $\beta$-strength function
is absent. The combined analysis of integral $\beta$-decay
characteristics: the half-lives and $\beta x n$ emission
probabilities ($P_{xn}$) helps to reconstruct the $\beta$-strength
function. A ratio of $P_{2n}/P_{1n}$ is a sensitive marker of the GT
strength distribution in continuum. This carries back the information
on the spin-isospin dependent components of the EDF. The
main aim of the present paper is to microscopically describe
the change of the $\beta$-strength function profile caused by the
$2p-2h$ fragmentation and to analyze its impact on the
$\beta$-decay half-lives and $\beta x n$-emission rates in
medium-heavy even-even Cd isotopes close to $N=82$ closed shell.

This paper is organized as follows. In Sec.~II, we apply the FRSA
model for studying the impact of the
PPC effects on the delayed multi-neutron emission. In Sec.~III we
describe the important ingredients used in
the $P_{xn}$ calculations and, in particular, $Q_{\beta}$-value,
the low-energy $2^+$ excitations of the parent nucleus, one and
two neutron separation energies for the daughter nucleus. We
analyze the results of the calculations of $\beta$-decay
half-lives in Sec.~IV A and the prediction of the $\beta x
n$-emission probabilities in Sec.~IV B.
Conclusions are finally drawn in Sec.~VI.
%
%===============================================================
%
\section{$\beta$-decay characteristics within the FRSA model}
The FRSA model for the charge-exchange excitations and the
$\beta$-decay was already introduced in Refs.~\cite{svg12,ss13}
and in Ref.~\cite{svbag14,e15}, respectively. In the present study
of the $\beta$-decay of even-even nuclei, this method is applied
for the prediction of the $\beta x n$ emission probabilities.

The $\beta x n$ emission is a multistep process consisting of
(a) the $\beta$-decay of the parent nucleus
(N, Z) which results in feeding the excited states of the daughter
nucleus (N - 1, Z + 1) followed by the (b)
$\beta x n$ emissions to the ground state and/or
(c) $\gamma$-deexcitation to the ground state of the product
nucleus (N - 1 - X, Z + 1). The starting point is the HF-BCS
calculation~\cite{RingSchuck} of the ground state within a
spherical symmetry assumption. The continuous part of the
single-particle spectrum is discretized by diagonalizing the HF
Hamiltonian on a harmonic oscillator basis. In the particle-hole
(p-h) channel we use the Skyrme interaction with the tensor
components and their inclusion leads to the modification of the
spin-orbit potential ~\cite{sbf77,c07,t43}. The pairing
correlations are generated by the density-dependent zero-range
force
\begin{equation}
\label{pair} V_{pair}({\bf r}_1,{\bf r}_2)=V_{0}\left(
1-\eta\left(\frac{\rho \left( r_{1}\right) } {\rho
_{0}}\right)^{\gamma}\right) \delta \left( {\bf r}_{1}-{\bf
r}_{2}\right),
\end{equation}
where $\rho _{0}$ is the nuclear saturation density. The values of
$V_{0}$, $\eta$ and $\gamma$ are fixed to reproduce the odd-even
mass difference of the studied nuclei~\cite{svg08,k90}. To
calculate binding energies of the daughter nucleus $B(N - 1, Z + 1)$
and the final nucleus $B(N - 1 - X, Z + 1)$, the blocking of the BCS
ground states~\cite{RingSchuck,block} is taken
into account. Finally, the calculated $Q_{\beta}$-value and the
neutron separation energies are given by
\begin{equation}
 Q_{\beta}=\Delta{M}_{n-H}+B(Z+1,N-1)-B(Z,N),
\end{equation}
\begin{equation}
 S_{xn}=B(Z+1,N-1)-B(Z+1,N-1-X).
\end{equation}
$\Delta M_{n-H}=0.782$~MeV is the mass difference between the neutron
and the hydrogen atom.

Constructing the QRPA equations on the
basis of HF-BCS quasiparticle states of the parent (even-even)
nucleus (N, Z) is the standard procedure~\cite{t05}.
The residual interactions in the p-h channel and the particle-particle channel
are derived consistently from the Skyrme EDF. The eigenvalues of
the QRPA equations are found numerically as the roots of
the FRSA secular equation for the cases of electric
excitations~\cite{gsv98,svg08} and charge-exchange
excitations~\cite{svg12,ss13}. It enables us to perform QRPA calculations in very large
two-quasiparticle (2QP) spaces. In particular, the cut-off of the
discretized continuous part of the single-particle spectra is
performed at the energy of 100~MeV. This is sufficient for exhausting
the Ikeda sum rule $S_{-}-S_{+}=3(N-Z)$.
A rather complete list of FRSA features can be found in Ref.~\cite{svbag14}.

To take into account the phonon-phonon coupling (PPC) effects we
follow the basic QPM ideas~\cite{solo,KS84}. The Hamiltonian can be
diagonalized in a space spanned by states composed of one and
two QRPA phonons~\cite{svbag14},
\begin{eqnarray}
\Psi _\nu (J M) = \left(\sum_iR_i(J \nu )Q_{J M i}^{+}\right.
\nonumber\\
\left.+\sum_{\lambda _1i_1\lambda _2i_2}P_{\lambda _2i_2}^{\lambda
_1i_1}( J \nu )\left[ Q_{\lambda _1\mu _1i_1}^{+}\bar{Q}_{\lambda
_2\mu _2i_2}^{+}\right] _{J M }\right)|0\rangle~, \label{wf}
\end{eqnarray}
where $\lambda$ denotes the total angular momentum and $\mu$ is
its z-projection in the laboratory system. The ground state of the
parent nucleus (N, Z) is the QRPA phonon vacuum
$\mid0\rangle$. The wave functions $Q_{\lambda \mu
i}^{+}\mid0\rangle$ of the one-phonon excited states of the
daughter nucleus (N - 1, Z + 1) are described as linear combinations of
2QP configurations; $\bar{Q}_{\lambda\mu i}^{+} |0\rangle$ is a
one-phonon electric excitation of the parent nucleus (N, Z).
The normalization condition for the wave functions~(\ref{wf}) is
\begin{equation}
\sum\limits_iR_i^2( J \nu)+ \sum_{\lambda _1i_1 \lambda _2i_2}
(P_{\lambda _2i_2}^{\lambda _1i_1}(J \nu))^2=1.
\end{equation}
For the unknown amplitudes $R_i(J\nu)$ and
$P_{\lambda_2i_2}^{\lambda_1i_1}(J\nu)$ the variational principle
leads to the set of linear equations with the rank equal to the
number of one- and two-phonon configurations, and for its solution
it is required to compute the Hamiltonian matrix elements coupling
one- and two-phonon configurations~\cite{svbag14,svbg13}. The
equations have the same form as the canonical QPM
equations~\cite{solo,KS84}, where the single-particle spectrum and
the residual interaction are derived from the same Skyrme EDF.

In the allowed GT approximation, the $\beta^{-}$-decay rate is
expressed by summing up the probabilities (in units of
$G_{A}^{2}/4\pi$) of the energetically allowed transitions
($E_{k}^{GT}\leq Q_{\beta}$) weighted with the integrated Fermi
function
\begin{eqnarray}
T_{1/2}^{-1}=\sum_k \lambda^{k}_{if}=D^{-1}\left(\frac{G_{A}}{G_{V}}\right)^{2}\nonumber\\
\times\sum\limits_{k}f_{0}(Z+1,A,E_{k}^{GT})B(GT)_{k},
\end{eqnarray}
\begin{equation}
 E_{k}^{GT}=Q_{\beta}-E_{1^+_k},
\end{equation}
where $\lambda^{k}_{if}$ is the partial $\beta^{-}$-decay rate,
$G_A/G_V$=1.25 is the ratio of the weak axial-vector and vector coupling
constants and $D$=6147 s (see Ref.~\cite{Suhonen}). $E_{1_k^+}$ denotes
the excitation energy of the daughter nucleus. As proposed in
Ref.~\cite{ebnds99}, this energy can be estimated by the following
expression:
\begin{equation}
E_{1^{+}_{k}}\approx E_{k}-E_{\textrm{2QP},\textrm{lowest}},
\end{equation}
where $E_{k}$ are the $1_k^+$ eigenvalues of the QRPA equations, or of
the equations taking into account the two-phonon
configurations~(\ref{wf}), and $E_{\textrm{2QP},\textrm{lowest}}$
corresponds the lowest 2QP energy. The
spin-parity of the lowest 2QP state is, in general, different from
$1^+$. The wave functions allow us to determine GT transitions whose
operator is $\hat{O}_{-} = \sum_{i, m} t_{-}(i) \sigma_m(i)$.
\begin{equation}
B(GT)_{k} = \left|\langle N- 1, Z+ 1; 1_k^+
|\hat{O}^{-}| N,Z; 0_{gs}^+ \rangle\right|^2 .
\end{equation}
Because of taking into account the tensor correlation effects
within the $1p-1h$ and $2p-2h$ configurational spaces,
 any quenching factors are redundant~\cite{bh82}.

The difference in the characteristic time scales of the $\beta$
decay and subsequent neutron emission processes justifies
an assumption of their statistical independence.
As proposed in Ref.~\cite{ps72}, the $P_{xn}$ probability of the
$\beta x n$ emission accompanying
the $\beta$ decay to the excited states in the daughter nucleus
can be expressed as
\begin{eqnarray}
P_{xn}=T_{1/2}D^{-1}\left(\frac{G_{A}}{G_{V}}\right)^{2}\nonumber\\
\times\sum\limits_{k'}f_{0}(Z+1,A,E_{k'}^{GT})B(GT)_{k'},
\end{eqnarray}
where the GT transition energy ($E_{k'}^{GT}$)
is located within the neutron emission window $Q_{\beta xn}\equiv
Q_{\beta}-S_{xn}$. For $P_{1n}$ we have $Q_{\beta 2n} \leq
E_{k'}^{GT} \leq Q_{\beta n}$, while for $P_{xn}$ this
becomes $Q_{\beta xn}{\leq}E_{k'}^{GT}{\leq}Q_{\beta x-1n}$.
Since we neglect the $\gamma$-deexcitation of the daughter
nucleus, some overestimation of the resulting $P_{xn}$ values can
be obtained~\cite{pn05}. The study of the $\gamma$-deexcitation
influence on the $P_{xn}$ values within our approach is in
progress.
%
%===============================================================
%
\section{Details of calculations}
\begin{figure}[t!]
\includegraphics[width=1.0\columnwidth]{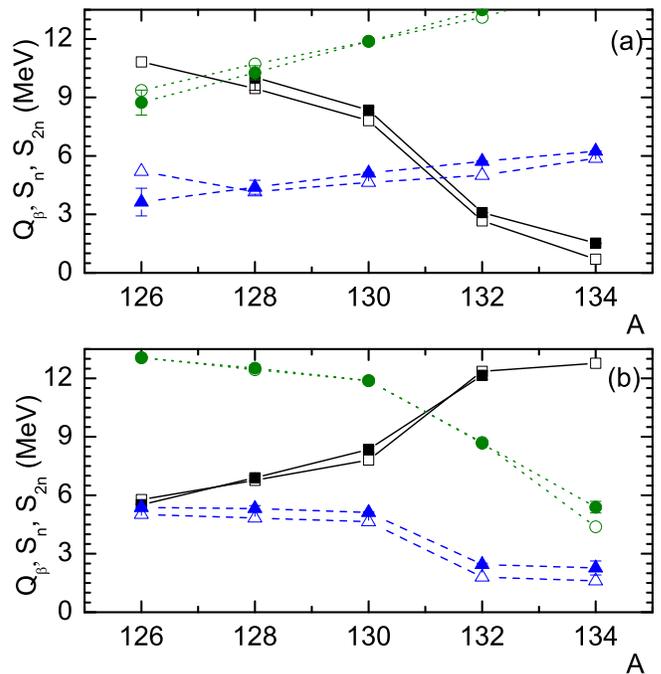}
\caption{(color online) Calculated $\beta$-decay windows
$Q_{\beta}$ of the parent nuclei
(squares), one- (triangles) and two- (circles) neutron separation
energies for the daughter nuclei. The upper
and lower panels correspond to the
even $N=82$ isotones and the even Cd isotopes, respectively.
Results of the HF-BCS calculations are denoted by the open
symbols. Experimental data (filled symbols) are taken from
Ref.~\cite{ame2012}.}
\end{figure}
\begin{figure}[t!]
\includegraphics[width=1.0\columnwidth]{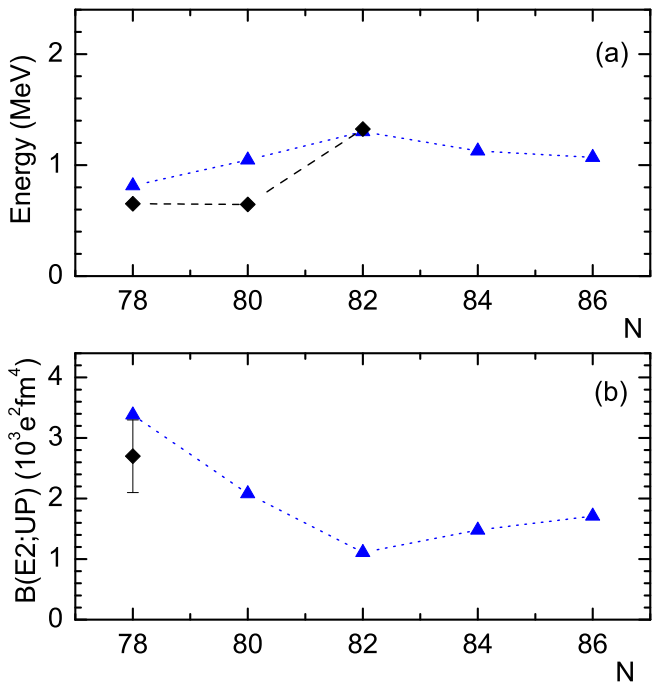}
\caption{(color online) Energies and $B(E2)$ values for
up-transitions to the $[2_1^+]_{QRPA}$ states in the
neutron-rich Cd isotopes. Results of the QRPA calculations
are denoted by the triangles. Experimental data (diamonds)
are taken from
Refs.~\cite{Ilieva14,Kautzsch00,Jungclaus07}.}
\end{figure}
\begin{figure}[t!]
\includegraphics[width=1.0\columnwidth]{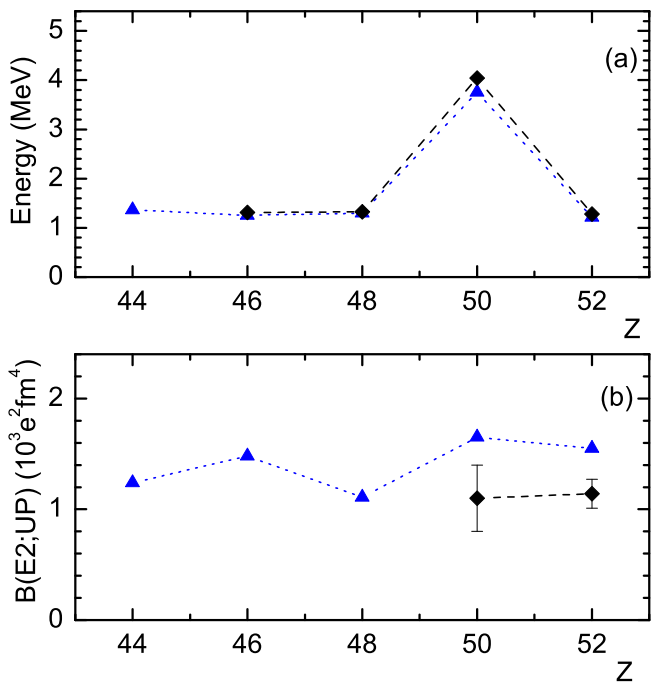}
\caption{Same as Fig. 2, but for the neutron-rich
$N=82$ isotones. Experimental data (diamonds) are taken from
Refs.~\cite{Danchev11,Varner05,Jungclaus07,Watanabe13}.}
\end{figure}
\begin{figure*}[t!]
\includegraphics[width=1.4\columnwidth]{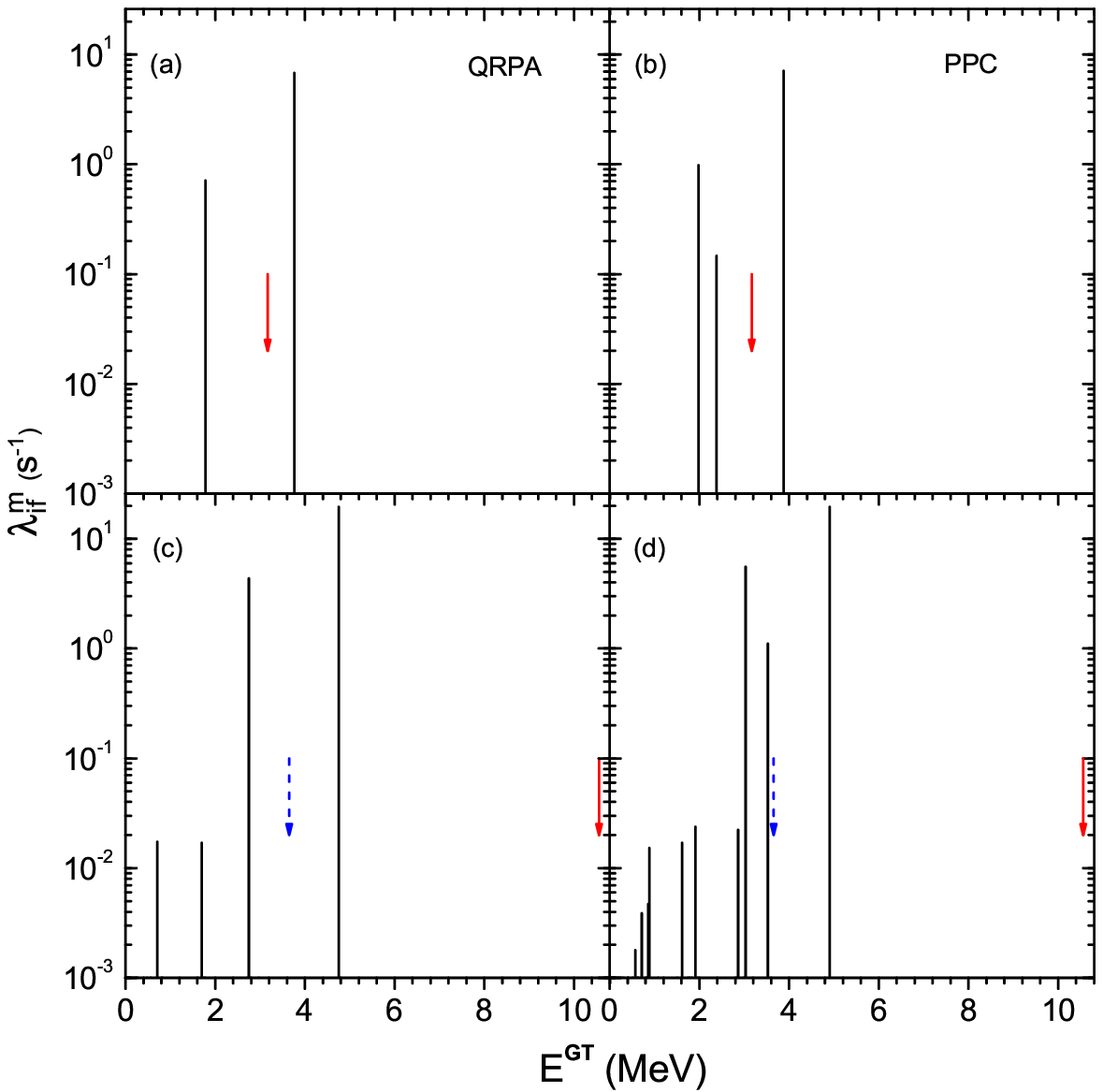}
\caption{(color online) The phonon-phonon coupling effect on the
$\beta$-transition rates in $^{130}$Cd (the upper panels) and
$^{132}$Cd (the lower panels). The left and right panels correspond
to the calculations within the  QRPA and taking into account
the $[1^{+}_{i}\otimes 2^{+}_{i'}]_{QRPA}$ configurations, respectively.
The calculated $Q_{\beta 1n}$ and $Q_{\beta 2n}$ energies are
denoted by the solid and dashed arrows, respectively.}
\end{figure*}
As the parameter set in the particle-hole channel,  we use
the Skyrme EDF T43 which takes into account the tensor term~\cite{t43}.
The T43 set is one of 36 parametrizations, covering a wide range of
the parameter space of the
isoscalar and isovector tensor term added with refitting the parameters
of the central interaction, where a fit protocol is very similar to that
of the successful SLy parametrizations. The spin-isospin Landau parameter
is given by
\begin{equation}
G_0^{\prime}=-N_0\left[ \frac{1}{4}t_0+\frac{1}{24}t_3\rho^{\alpha_3}+\frac{1}{8}k^2_F(t_1-t_2)\right],
\end{equation}
where $N_0 = 2k_Fm^{*}/\pi^2\hbar^2$ is the level density, with
$k_F$ being the Fermi momentum and $m^{*}$ the nucleon effective
mass. At saturation density ($\rho=\rho_0$), the set T43 predicts
enough positive value for $G_0^{\prime}= 0.14$ and it gives a
reasonable description of properties of the GT and charge-exchange
spin-dipole resonances~\cite{sag11}. Using the PPC effects within
the FRSA model, the T43 set gives a reasonable agreement with
experimental data for the $\beta$-decay half-life of the
neutron-rich doubly-magic nucleus $^{132}$Sn, see Sec.~IV A. It is
worth mentioning that the first study of the strong impact of the
tensor correlations on the $^{132}$Sn half-life has been done
in~\cite{mb13}.

The pairing correlations are generated by a surface-peaked pairing
force~(\ref{pair}) with $\eta=$1, $\gamma$=1 and the value
$\rho_{0}$= 0.16fm~$^{-3}$ for the nuclear saturation density.
Using the soft cutoff at 10~MeV above the Fermi energies,
the pairing strength is fixed to be $V_{0}$=$-870$~MeV~fm$^3$
in order to fit  the experimental neutron pairing gaps of
$^{126,128}$Cd, $^{130}$Sn, $^{132}$Te obtained by
the three-point formula~\cite{svg12,ss13}.

The correct description of the $Q_\beta$ values for the
parent nuclei and the neutron separation
energies ($S_{xn}$) for the daughter nuclei
is the important ingredient for the reliable prediction of the
$P_{2n}/P_{1n}$
ratios. The binding energies of the daughter and final nuclei are
calculated with the blocking effect for unpaired
nucleons~\cite{RingSchuck,block}. For $^{126}$Rh, $^{128}$Ag,
$^{126,128,130}$In, $^{132}$Sb and $^{134}$I,  the neutron
quasiparticle blocking is based on filling the $1h_{11/2}$
subshell and the $2f_{7/2}$ subshell should be blocked for
$^{132,134}$In. The proton $1g_{9/2}$ and $1g_{7/2}$ subshells are
chosen to be blocked in the cases of the Rh, Ag, In isotopes  and
the Sb, I isotopes, respectively. The calculated $Q_{\beta}$,
$S_{n}$ and $S_{2n}$ values of the $\beta$-decays of the even Cd
isotopes and the  $N=82$ isotones are compared with the
experimental data~\cite{ame2012} in Fig.1. The existing experimental data
show a different $A$-behavior, namely, the 6.6-time
reduction of $Q_{\beta}$  values from $^{128}$Pd to $^{134}$Te and
the gradual reduction of $Q_{\beta}$ values with decreasing
neutron number for the Cd isotopes. The results of the HF-BCS
calculation with the T43 EDF are in a reasonable agreement with
the experimental data.

In order to construct the wave
functions~(\ref{wf}) of the low-energy $1^{+}$ states, in the
present study we assume only the $[1^{+}_{i}\otimes
2^{+}_{i'}]_{QRPA}$ terms
for separating the sole impact of the quadrupole-phonon
coupling. All one- and two-phonon configurations with the
excitation energy of the daughter nucleus $E_{1_k^+}$ up to 16~MeV
are included. We have checked that the inclusion of the
high-energy configurations leads to minor effects on the half-life
values. As it is pointed out in Ref.~\cite{svbag14}, the
$[1_1^+\otimes2_1^+]_{QRPA}$ configuration is the important
ingredient for the half-life description since the
$[2_1^+]_{QRPA}$ state is the lowest collective excitation which
leads to the minimal two-phonon energy and the maximal Hamiltonian
matrix elements for coupling of the one- and two-phonon configurations.

It is interesting to examine the energies and transition
probabilities of the $[2_1^+]_{QRPA}$ states of the neutron-rich
Cd isotopes (see Fig.2). There is a significant increase of the
$2_1^+$ energy of $^{130}$Cd. It
corresponds to a standard evolution of the $2_1^+$ energy near
closed shells. In all five nuclei, the $2^+_1$ wave functions are
dominated by the proton configuration $\{1g_{9/2},1g_{9/2}\}_\pi$
($>$73\%). The closure of the neutron subshell $1h_{11/2}$ in
$^{130}$Cd, leads to the vanishing of the neutron pairing and as a
result the lowest neutron 2QP energy $\{2f_{7/2},1h_{11/2}\}_\nu$
is larger than the lowest neutron 2QP energies
$\{1h_{11/2},1h_{11/2}\}_\nu$ in $^{128}$Cd and
$\{2f_{7/2},2f_{7/2}\}_\nu$ in $^{132}$Cd. Correspondingly the
$2_1^{+}$ state of $^{130}$Cd has
noncollective structure with the $\{1g_{9/2},1g_{9/2}\}_\pi$
domination (about 96\%) and the $B(E2)$ value is reduced.
As the data for the Cd isotopes are very scarce, the $N=82$ isotones
are used for reference. Results of
QRPA calculation with T43 EDF and the experimental
data~\cite{Danchev11,Varner05,Jungclaus07,Watanabe13} are shown in
Fig.~3. The calculated values are in a reasonable agreement with
the data. Mooving along the neutron-rich
$N=82$ isotones chain one can find  that $2_1^+$ states in
$^{126}$Ru, $^{128}$Pd and $^{130}$Cd as is discussed above,  have
a noncollective structure with a domination of the
$\{1g_{9/2},1g_{9/2}\}_\pi$ configuration. In $^{132}$Sn, the main
configurations are the neutron $\{2f_{7/2},1h_{11/2}\}_\nu$ (56\%)
and the proton $\{2d_{5/2},1g_{9/2}\}_\pi$ (37\%) ones. In
$^{134}$Te, the $2_{1}^+$ state is dominated by the lowest 2QP
component $\{1g_{7/2},1g_{7/2}\}_\pi$. The structure peculiarities
are reflected in the $B(E2)$ behavior in this chain. We find a
satisfactory description of the isotonic dependence of the $2_1^+$
energy  near the closed proton shell.
%
%===============================================================
%
\section{Results}
Using the same set of parameters the main features of the
$\beta$-decay and available $\beta x
n$ rates are described for the neutron-rich nuclei
$^{132}_{\,\,\,50}$Sn,
$^{126,128,130,132,134}_{\,\,\,\,\,\,\,\,\,\,\,\,\,\,\,\,\,\,\,\,\,\,\,\,\,\,\,\,\,\,\,\,\,\,\,\,\,\,\,\,\,\,\,\,\,48}$Cd,
$^{128}_{\,\,\,46}$Pd and $^{126}_{\,\,\,44}$Ru. The integral
$\beta$-decay observables are substantially defined by the
structure of the $\beta$-strength function. Fig.4 depicts the
$\beta$-strength function of $^{130,132}$Cd (in terms of the
transition rate) calculated a), c) within the QRPA and
b), d) with the $[1^{+}_{i}\otimes 2^{+}_{i'}]_{QRPA}$
configurations taken into account.

For $^{130}$Cd, the QRPA strength function has a rather simple
two-peak structure, the main transition to the $[1^+_1]_{QRPA}$
state is built on the $\{\pi1g_{9/2},\nu1g_{7/2}\}$ configuration.
Inclusion of the PPC shifts main peak by +120 keV increasing its
amplitude by 5\%  and also an additional low rate peak at $
E_{2}^{GT}=$2.4~MeV comes from the $[1_1^+\otimes2_1^+]_{QRPA}$
configuration (93\%). Thus an additional peak is dominated by
the  four-quasiparticle configuration
$\{\pi1g_{9/2},\pi1g_{9/2},\pi1g_{9/2},\nu1g_{7/2}\}$. On the
other hand, the main contribution to the GT matrix element comes
from the one-phonon configuration $[1^+_1]_{QRPA}$ which exhausts
about 6\% of the $1^+_2$ wave function. These changes are
translated into the corresponding half-life reduction and
$P_{tot}$ growth. This will be discussed in the next subsections.
As can be seen from Fig.4, we get the similar tendency of the QRPA
strength function in the case of $^{132}$Cd. However there are remarkable
changes in the values of the $Q_{\beta 1n}$ and $Q_{\beta 2n}$
windows. It is seen that the calculated
neutron-emission probability ($P_{tot}$) exhausts 100\% since
all the GT transition energies are less than
$Q_{\beta 1n}$. The additional (two-phonon) peak leads to the
$P_{2n}/P_{1n}$ increase, see Sec.~IV B.
\subsection{$\beta$-decay half-lives}
\begin{table}
\caption{
The phonon-phonon coupling effect on $\beta$-decay
half-lives of the neutron-rich Cd isotopes.
Experimental data are taken from~\cite{L15}.}
\begin{ruledtabular}
\begin{tabular}{ccccc}
        &\multicolumn{3}{c}{Half-life (ms)}\\
Nucleus & QRPA & PPC & Expt.          \\
\noalign{\smallskip}\hline\noalign{\smallskip}$^{126}$Cd
        & 334  & 263 & $513{\pm}6$    \\
\noalign{\smallskip}\noalign{\smallskip}$^{128}$Cd
        & 212  & 180 & $245{\pm}5$    \\
\noalign{\smallskip}\noalign{\smallskip}$^{130}$Cd
        & 133  & 121 & $127{\pm}2$    \\
\noalign{\smallskip}\noalign{\smallskip}$^{132}$Cd
        &  42  &  38 &  $82{\pm}4$    \\
\noalign{\smallskip}\noalign{\smallskip}$^{134}$Cd
        &  36  &  32 &  $65{\pm}15$    \\
\end{tabular}
\end{ruledtabular}
\end{table}
\begin{figure}[t!]
\includegraphics[width=1.0\columnwidth]{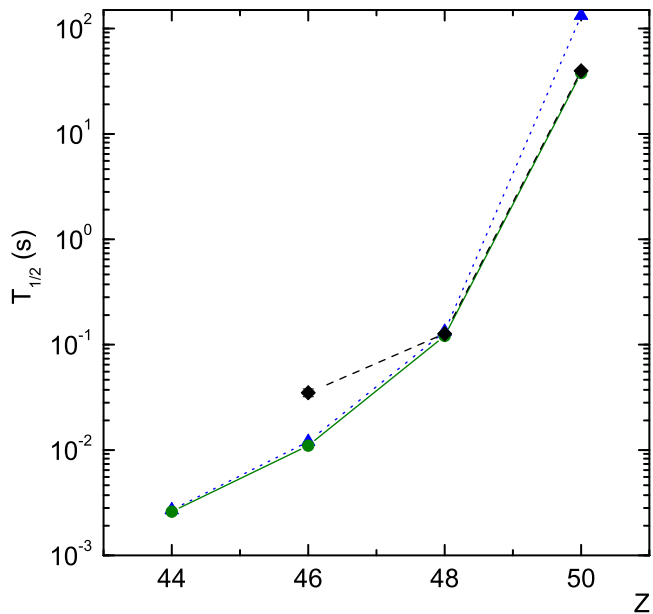}
\caption{(color online) The phonon-phonon coupling effect on $\beta$-decay
half-lives of the neutron-rich $N=82$ isotones. The circles correspond to the
half-lives calculated with inclusion of the $[1^{+}_{i}\otimes 2^{+}_{i'}]_{QRPA}$
configurations, the triangles are the QRPA calculations. Experimental
data (diamonds) are from Refs.~\cite{Mach95,L15}.}
\end{figure}
For $^{126,128,130,132,134}$Cd, the results of our
calculations and experimental data~\cite{L15} are shown in Table~1.
First, the half-lives are studied within the one-phonon
approximation. At a qualitative level, our results reproduce the
experimental mass dependence of the
half-lives. The largest contribution ($>$80\%) in
all the calculated half-lives comes from the $[1_1^+]_{QRPA}$ state.
The dominant configuration of the $[1_1^+]_{QRPA}$ states is
$\{\pi1g_{9/2},\nu1g_{7/2}\}$ with the
contribution of about 99~$\%$, and $\log ft
\approx 2.9$ in all the Cd isotopes considered. Let us consider how to explain the
3.2-time reduction of calculated half-life values from $^{130}$Cd
to $^{132}$Cd, see Table 1. The $\{\pi1g_{9/2},\nu1g_{7/2}\}$
energy is equal to 7.5~MeV for $^{130}$Cd and 9.5~MeV for
$^{132}$Cd. Also we find that the lowest 2QP energy is  either the
$\{\pi1g_{9/2},\nu1h_{11/2}\}$ value of 3.4~MeV for $^{130}$Cd, or
$\{\pi1g_{9/2},\nu2f_{7/2}\}$ value of 1.9~MeV for $^{132}$Cd. As
a result, the excitation energy of the first $1^+$ state is
increased from $^{130}$In to $^{132}$In. Therefore, the 4.6~MeV
increase of the $Q_{\beta}$ values (see Fig.~1) plays the key role
to explain this half-life reduction. The
analysis within the one-phonon approximation can help to
explain the main peculiarities of the half-lives
A-dependence, but it is only a rough estimate.

Let us now discuss the extension of the space to one- and
two-phonon configurations on the half-lives. As expected, the
largest contribution ($>$70\%) in half-life comes from the $1_1^+$
state calculated with the PPC. The dominant contribution in the
wave function of the first $1^+$ state comes from the
$[1^+_1]_{QRPA}$ configuration, but the
$[1_1^+\otimes2_1^+]_{QRPA}$ contribution is appreciable.
Inclusion of the two-phonon terms results in a decrease of the
$1^+_1$ energy. The well-known experimental characteristics
of the $1^+_1$ state in $^{130}$In is a stringent test for
the existing microscopic approaches~\cite{C07}. Results of FRSA
model with the T43 EDF ($E_{1^{+}_{1}}=3.9$~MeV and $\log ft =
3.0$) can be compared with the experimental excitation energy
$E_{1^{+}_{1}}=2.12$~MeV and $\log ft = 4.1$~\cite{D03}. The
calculated two-quasiparticle energy and unperturbed B(GT) value
are too large to be properly renormalized the inclusion of the two-phonon
configurations. One may possibly seek for improvements of
the $T=0$ pairing term in the EDF used. Table 1 shows the
half-life reduction as an effect of the quadrupole-phonon
coupling, see, e.g., Fig.~4. The calculated half-life of
$^{130}$Cd is in excellent agreement with the experimental
data~\cite{L15}. The discrepancies between measured and calculated
half-lives of $^{126,132,134}$Cd are due to total neglect of the
first-forbidden transitions.

It is worth to mention that the EDF T43 within the QRPA gives a
satisfactory agreement with experimental data for the
$\beta$-decay half-life of the neutron-rich doubly-magic nucleus
$^{132}$Sn~\cite{mb13}. The PPC effect results in a improvement of
the half-life description, see Fig.~5. As can be seen from Table~1
and Fig.~5, there are different behavior of the existing experimental
half-lives~\cite{Mach95,L15}, namely, the 313-time reduction of
half-life values from $^{132}_{\,\,\,50}$Sn to
$^{130}_{\,\,\,48}$Cd and the gradual reduction of half-lives with
increasing neutron number for $^{126,128,130}$Cd. One can see that
FRSA model with the T43 EDF reproduces this behaviour and our
results predict the half-lives for $^{128}_{\,\,\,46}$Pd and
$^{126}_{\,\,\,44}$Ru. Further
an improvement can be achieved if the first-forbidden
transitions are taken into account \cite{ff03}.
It is planned to extend our formalism including the
first-forbidden transitions.
%
%===============================================================
%
\subsection{Probabilities of the $\beta$-delayed neutron emission}
\begin{figure}[t!]
\includegraphics[width=1.0\columnwidth]{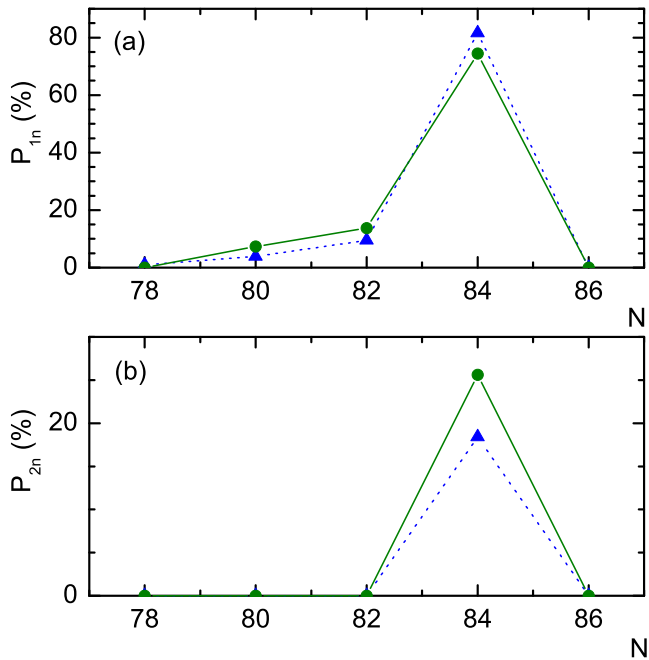}
\caption{(color online) The phonon-phonon coupling effect on $\beta$-delayed neutron-emission
probabilities of the neutron-rich Cd isotopes. The circles correspond to the
probabilities calculated with inclusion of the $[1^{+}_{i}\otimes 2^{+}_{i'}]_{QRPA}$
configurations, the triangles are the QRPA calculations.}
\end{figure}
An additional constraints on the $\beta$-strength function are
given by the total and multi-neutron emission probabilities.
The well-known experimental probabilities are $P_{tot}=3.5\pm1.0$\%
for $^{130}$Cd~\cite{D03} and $60\pm15\%$ for $^{132}$Cd~\cite{H01}.
The calculated $P_{1n,2n}$ values are displayed in Fig.~6.
For $^{126}$Cd, one has to mention a non-zero $P_{tot}$ value less
than 1\%. For $^{130}$Cd, the calculated $P_{tot}$ value of 13.7\%
is higher than the experimental value of 3.5\%~\cite{D03} which
may indicate the necessity of including the $T=0$
pairing interaction. For all considered isotopes we obtain the maximal
$P_{1n}$ and $P_{2n}$ values in the case of $^{132}$Cd.
The redistribution of the QRPA
estimate in favor of $P_{2n}$ occurs when the PPC effect is
included: $P_{2n}$= 25.6\% compared to  $P_{1n}$= 74.4\%. This
differs from the DF3a+cQRPA prediction by Ref.~\cite{B14} which
gives $P_{1n}$=84.13\%, $P_{2n}$=0.14\%. Also, it is interesting
to compare the $P_{2n}/P_{1n}$ ratios. They are 0.002 in
Ref.~\cite{B14} and 0.34 in the present calculation. The
difference between the predictions of two models can be explained
by the sensitivity of multi-neutron delayed emission to the
details of  the $\beta$-strength function and the neutron
separation energies. Also further from
the closed shell ($^{134}$Cd) one can not neglect the increasing
contribution of the first-forbidden transition known to reduce the
$P_{tot}$ value~\cite{pn05,B08,B14} and substantially
redistributing $P_{xn}$ values. As is shown in Ref.~\cite{B14},
the DF3a+cQRPA calculation gives $P_{1n}$=32.75\%,
$P_{2n}$=47.47\% and $P_{3n}$=7.12\%.

It would be instructive to study the PPC effect on the $P_{1n,2n}$ values of the
neutron-rich nuclei. For $^{126,128,130}$Cd within the QRPA there are two main GT decays which
define the $P_{tot}$ values (see for example, Fig.~4). The inclusion of the PPC has
a stronger influence on the energy shift
of the $1^+_3$ state which is mainly built on the $[1^+_2]_{QRPA}$ configuration. Also the
two-phonon state $1^+_2$ appears. Both effects increase the $P_{tot}$ value.
In the case of $^{132}$Cd these dynamic features of $1^+_2$ and $1^+_3$ states are responsible for
increasing the $P_{2n}$ value. The QRPA value $P_{2n}/P_{1n}$=0.22 is replaced by 0.34 with the
PPC included.  Notice that $P_{2n}/P_{1n}$=0.29 if the $1^+_2$ state is not taken into account.
%
%=====================================================================
%
%
\section{Conclusions}
Starting from the Skyrme mean-field calculations, we have studied
the effects of the phonon-phonon coupling on the properties of the
$\beta$-delayed multi-neutron emission and, in particular, on
$P_{2n}/P_{1n}$ ratios of nuclei in the
mass range $A\approx130$. The finite-rank separable approach
to the QRPA problem enables one to
perform the calculations in very large configurational spaces.

The parametrization T43 of the Skyrme interaction is used for all
calculations in connection with the surface-peaked zero-range
pairing interaction.
In particular, we study the multi-neutron emission in
the $^{132}$Cd in comparison with $N=82$ isotone $^{130}$Cd.
We have found a significant  two-neutron emission for $^{132}$Cd,
the effect which  was predicted within the FRDM+QRPA and the
DF3+cQRPA before. Notice, that as well as in the DF3+cQRPA calculations,
our results from the Skyrme interaction are in reasonable agreement with
experimental half-lives. It is the first successful description obtained
with the Skyrme interaction for the experimental neutron-emission probabilities.
The coupling between
one- and two-phonon terms in the wave functions of $1^{+}$ states
is shown to be essential. The QRPA underestimates the $P_{2n}/P_{1n}$
values. Inclusion of the two-phonon configurations produces
an impact on the $P_{2n}$ value which leads to the 55\% increase of
the $P_{2n}/P_{1n}$ ratio. For $^{126,128,130,132,134}$Cd, the maximal
$P_{1n}$ and $P_{2n}$ values are obtained in the case of
$^{132}$Cd. For $^{126}$Cd, a nonzero probability of
the neutron emission is found.

We conclude that the present approach makes it possible to perform
the new microscopic analysis of the rates of the $\beta$-delayed
multi-neutron emission. The model can be extended by enlarging the
variational space for the $1^{+}$ states with the inclusion of the
two-phonon configurations constructed from phonons with monopole,
dipole and octupole multipolarities. The computational
developments that would allow us to conclude on this point are
underway.
%==================================================================
%
\section*{Acknowledgments}
We thank Nguyen Van Giai, Yu.E. Penionzhkevich, H. Sagawa and D.
Verney for useful discussions.  This work is
supported by the Russian Science Foundation
(Grant No. RSF-16-12-10161).
%
%==================================================================
%


\begin{thebibliography}{99}
\bibitem{LM03}
K. Langanke, and G. Mart\'{i}nez-Pinedo, Rev. Mod. Phys. {\bf 75},
819 (2003).
\bibitem{G60}
V. I. Goldansky, Nucl. Phys. {\bf 19}, 482 (1960).
\bibitem{A79}
R. E. Azuma, L. C. Carraz, P. G. Hansen, B. Jonson, K.-L. Kratz,
S. Mattsson, G. Nyman, H. Ohm, H. L. Ravn, A. Schr\"{o}der, and W.
Ziegert, Phys. Rev. Lett. {\bf 43}, 1652 (1979).
\bibitem{D80}
C. Detraz, M. Epherre, D. Guillemaud, P. G. Hansen, B. Jonson, R.
Klapisch, M. Langevin, S. Mattsson, F. Naulin, G. Nyman, A. M.
Poskanzer, H. L. Ravn, M. de Saint-Simon, K. Takahashi, C.
Thibault, and F. Touchard, Phys. Lett. B {\bf 94}, 307 (1980).
\bibitem{L85}
Yu. S. Lyutostansky, V. K. Sirotkin, and I. V. Panov, Phys. Lett.
B {\bf 161}, 9 (1985).
\bibitem{S12}
A. Spyrou, Z. Kohley, T. Baumann, D. Bazin, B. A. Brown, G.
Christian, P. A. DeYoung, J. E. Finck, N. Frank, E. Lunderberg, S.
Mosby, W. A. Peters, A. Schiller, J. K. Smith, J. Snyder, M. J.
Strongman, M. Thoennessen, and A. Volya, Phys. Rev. Lett. {\bf
108}, 102501 (2012).
\bibitem{M12}
F. M. Marqu\'{e}s, N. A. Orr, N. L. Achouri, F. Delaunay, and J.
Gibelin, Phys. Rev. Lett. {\bf 109}, 239201 (2012).
\bibitem{B16}
C. A. Bertulani, and V. Zelevinsky, Nature {\bf 532}, 448
(2016).
\bibitem{C16}
R. Caballero-Folch, C. Domingo-Pardo, J. Agramunt, A. Algora, F.
Ameil, A. Arcones, Y. Ayyad, J. Benlliure, I. N. Borzov, M. Bowry,
F. Calvi\~{n}o, D. Cano-Ott, G. Cort\'{e}s, T. Davinson, I.
Dillmann, A. Estrade, A. Evdokimov, T. Faestermann, F. Farinon, D.
Galaviz, A. R. Garc\'{\i}a, H. Geissel, W. Gelletly, R.
Gernh\"{a}user, M. B. G\'{o}mez-Hornillos, C. Guerrero, M. Heil,
C. Hinke, R. Kn\"{o}bel, I. Kojouharov, J. Kurcewicz, N. Kurz, Yu.
A. Litvinov, L. Maier, J. Marganiec, T. Marketin, M. Marta, T.
Mart\'{\i}nez, G. Mart\'{\i}nez-Pinedo, F. Montes, I. Mukha, D. R.
Napoli, C. Nociforo, C. Paradela, S. Pietri, Zs. Podoly\'{a}k, A.
Prochazka, S. Rice, A. Riego, B. Rubio, H. Schaffner, Ch.
Scheidenberger, K. Smith, E. Sokol, K. Steiger, B. Sun, J. L.
Ta\'{\i}n, M. Takechi, D. Testov, H. Weick, E. Wilson, J. S.
Winfield, R. Wood, P. Woods, and A. Yeremin, Phys. Rev. Lett. {\bf
117}, 012501 (2016).
\bibitem{ff03}
I. N. Borzov, Phys. Rev. C {\bf 67}, 025802 (2003).
\bibitem{pn05}
I.N. Borzov, Phys. Rev.C {\bf 71}, 065801 (2005).
\bibitem{w97}
T. Wakasa, H. Sakai, H. Okamura, H. Otsu, S. Fujita, S. Ishida, N. Sakamoto,
T. Uesaka, Y. Satou, M. B. Greenfield, and K. Hatanaka,
Phys. Rev. C {\bf 55}, 2909 (1997).
\bibitem{mda}
M. Ichimura, H. Sakai, and T. Wakasa, Prog. Part. Nucl. Phys. {\bf
56}, 446 (2006).
\bibitem{bh82}
G. F. Bertsch and I. Hamamoto,  Phys. Rev. C {\bf 26}, 1323
(1982).
\bibitem{KS84}
V. A. Kuzmin and V. G. Soloviev, J. Phys. {\bf G10}, 1507 (1984).
\bibitem{dosw87}
S. Drozdz, F. Osterfeld, J. Speth, and J. Wambach, Phys. Lett.
{\bf B189}, 271 (1987).
\bibitem{cnbb94}
G. Col\'o, Nguyen Van Giai, P. F. Bortignon, and R. A. Broglia,
Phys. Rev. C {\bf 50}, 1496 (1994).
\bibitem{csnbs98}
G. Col\'o, H. Sagawa, Nguyen Van Giai, P. F. Bortignon, and T.
Suzuki, Phys. Rev. C {\bf 57}, 3049 (1998).
\bibitem{ncbbm12}
Y. F. Niu, G. Col\`o, M. Brenna, P. F. Bortignon, and J. Meng,
Phys. Rev. C {\bf 85}, 034314 (2012).
\bibitem{solo}
 V. G. Soloviev, {\it Theory of Atomic Nuclei: Quasiparticles and
 Phonons} (Institute of Physics, Bristol and Philadelphia, 1992).
\bibitem{KS85}
V. A. Kuzmin and V. G. Soloviev, J. Phys. {\bf G11}, 603 (1985).
\bibitem{gsv98}
 Nguyen Van Giai, Ch. Stoyanov, and V. V. Voronov, Phys. Rev. C {\bf
 57}, 1204 (1998).
\bibitem{svg04}
 A. P. Severyukhin, V. V. Voronov, and Nguyen Van Giai, Eur. Phys.
 J. {\bf A22}, 397 (2004).
\bibitem{svg12}
 A. P. Severyukhin, V. V. Voronov, and Nguyen Van Giai, Prog. Theor.
 Phys. {\bf 128}, 489 (2012).
\bibitem{svbag14}
 A. P. Severyukhin, V. V. Voronov, I. N. Borzov, N. N. Arsenyev, and Nguyen Van Giai,
 Phys. Rev. C {\bf 90}, 044320 (2014).
\bibitem{H01}
M. Hannawald, V. N. Fedoseyev, U. K\"{o}ster, K.-L. Kratz, V. I.
Mishin, W. F. Mueller, H. L. Ravn, J. Van Roosbroeck, H. Schatz,
V. Sebastian, W. B. Walters, and the ISOLDE Collaboration, Nucl.
Phys. A {\bf 688}, 578c (2001).
\bibitem{D03}
I. Dillmann, K.-L. Kratz, A.W\"{o}hr, O. Arndt, B. A. Brown, P. Hoff, M. Hjorth-Jensen, U. K\"{o}ster,
A. N. Ostrowski, B. Pfeiffer, D. Seweryniak, J. Shergur, W. B.Walters, and the ISOLDE Collaboration,
Phys. Rev. Lett. {\bf 91}, 162503 (2003).
\bibitem{L15}
G. Lorusso, S. Nishimura, Z. Y. Xu, A. Jungclaus, Y. Shimizu, G.
S. Simpson, P.-A. S\"{o}derstr\"{o}m, H. Watanabe, F. Browne, P.
Doornenbal, G. Gey, H. S. Jung, B. Meyer, T. Sumikama, J.
Taprogge, Zs. Vajta, J. Wu, H. Baba,1 G. Benzoni, K. Y. Chae, F.
C. L. Crespi, N. Fukuda, R. Gernh\"{a}user, N. Inabe, T. Isobe, T.
Kajino, D. Kameda, G. D. Kim, Y.-K. Kim, I. Kojouharov, F. G.
Kondev, T. Kubo, N. Kurz, Y. K. Kwon, G. J. Lane, Z. Li, A.
Montaner-Piz\'{a}, K. Moschner, F. Naqvi, M. Niikura, H.
Nishibata, A. Odahara, R. Orlandi, Z. Patel, Zs. Podoly\'{a}k, H.
Sakurai, H. Schaffner, P. Schury, S. Shibagaki, K. Steiger, H.
Suzuki, H. Takeda, A. Wendt, A. Yagi, and K. Yoshinaga, Phys. Rev.
Lett. {\bf 114}, 192501 (2015).
\bibitem{D16}
R. Dunlop, V. Bildstein, I. Dillmann, A. Jungclaus, C. E. Svensson, C. Andreoiu, G. C. Ball, N. Bernier,
H. Bidaman, P. Boubel, C. Burbadge, R. Caballero-Folch, M. R. Dunlop, L. J. Evitts, F. Garcia, A. B. Garnsworthy,
P. E. Garrett, G. Hackman, S. Hallam, J. Henderson, S. Ilyushkin, D. Kisliuk, R. Kr\"{u}cken, J. Lassen, R. Li,
E. MacConnachie, A. D. MacLean, E. McGee, M. Moukaddam, B. Olaizola, E. Padilla-Rodal, J. Park, O. Paetkau,
C. M. Petrache, J. L. Pore, A. J. Radich, P. Ruotsalainen, J. Smallcombe, J. K. Smith, S. L. Tabor, A. Teigelh\"{o}fer,
J. Turko, and T. Zidar,
 Phys. Rev. C {\bf 93}, 062801(R) (2016).
\bibitem{J16}
A. Jungclaus, H. Grawe, S. Nishimura, P. Doornenbal, G. Lorusso, G. S. Simpson, P.-A. S\"{o}derstr\"{o}m, T. Sumikama,
J. Taprogge, Z. Y. Xu, H. Baba, F. Browne, N. Fukuda, R. Gernh\"{a}user, G. Gey, N. Inabe, T. Isobe,
H. S. Jung, D. Kameda, G. D. Kim, Y.-K. Kim, I. Kojouharov, T. Kubo, N. Kurz, Y. K. Kwon, Z. Li,
H. Sakurai, H. Schaffner, Y. Shimizu, K. Steiger, H. Suzuki, H. Takeda, Zs. Vajta, H. Watanabe, J. Wu,
A. Yagi, K. Yoshinaga, G. Benzoni, S. B\"{o}nig, K. Y. Chae, L. Coraggio, J.-M. Daugas, F. Drouet, A. Gadea,
A. Gargano, S. Ilieva, N. Itaco, F. G. Kondev, T. Kr\"{o}ll, G. J. Lane, A. Montaner-Piz\'{a}, K. Moschner,
D. M\"{u}cher, F. Naqvi, M. Niikura, H. Nishibata, A. Odahara, R. Orlandi, Z. Patel, Zs. Podoly\'{a}k, and A. Wendt,
 Phys. Rev. C {\bf 94}, 024303 (2016).
\bibitem{M97}
P. M\"{o}ller, J. R. Nix, and K.-L. Kratz, At. Data Nucl. Data
Tables {\bf 66}, 131 (1997).
\bibitem{pn06}
I.N. Borzov, Nucl.Phys. A {\bf 777}, 645 (2006).
\bibitem{B08}
I. N. Borzov, J. J. Cuenca-Garc\'{\i}a, K. Langanke,
G. Mart\'{\i}nez-Pinedo, and F. Montes, Nucl. Phys. A {\bf 814}, 159 (2008).
\bibitem{Marketin16}
T. Marketin, L. Huther, and G. Mart\'{i}nez-Pinedo, Phys. Rev. C
{\bf 93}, 025805 (2016).
\bibitem{Mustonen16}
M. T. Mustonen, and J. Engel, Phys. Rev. C {\bf 93}, 014304
(2016).
\bibitem{M03}
P. M\"{o}ller, B. Pfeiffer, and K.-L. Kratz, Phys. Rev. C {\bf
67}, 055802 (2003).
\bibitem{Z13}
Q. Zhi, E. Caurier, J. J. Cuenca-Garc\'{i}a, K. Langanke,  G.
Mart\'{i}nez-Pinedo, and K. Sieja, Phys. Rev. C {\bf 87}, 025803
(2013).
\bibitem{ss13}
 A. P. Severyukhin and H. Sagawa, Prog. Theor. Exp. Phys. {\bf 2013},
 103D03 (2013).
\bibitem{e15}
A. Etil\'{e}, D. Verney, N. N. Arsenyev, J. Bettane, I. N. Borzov, M. Cheikh Mhamed, P. V. Cuong, C. Delafosse,
F. Didierjean, C. Gaulard, Nguyen Van Giai, A. Goasduff, F. Ibrahim, K. Kolos, C. Lau, M. Niikura, S. Roccia,
A. P. Severyukhin, D. Testov, S. Tusseau-Nenez, and V. V. Voronov,
Phys. Rev. C {\bf 91}, 064317 (2015).
\bibitem{RingSchuck}
 P. Ring and P. Schuck, {\it The Nuclear Many Body Problem}
 (Springer, Berlin, 1980).
\bibitem{sbf77}
F. Stancu, D. M. Brink, and H. Flocard, Phys. Lett. {\bf B68}, 108 (1977).
\bibitem{c07}
G. Col\`o, H. Sagawa, S. Fracasso, and P.F. Bortignon, Phys. Lett. {\bf B646}, 227 (2007);
Phys. Lett. {\bf B668}, 457(E) (2008).
\bibitem{t43}
T. Lesinski, M. Bender, K. Bennaceur, T. Duguet, and J.Meyer,
Phys. Rev. C {\bf 76}, 014312 (2007).
\bibitem{svg08}
 A. P. Severyukhin, V. V. Voronov, and Nguyen Van Giai, Phys. Rev.
 C {\bf 77}, 024322 (2008).
\bibitem{k90}
S. J. Krieger, P. Bonche, H. Flocard, P. Quentin, and M. S. Weiss,
Nucl. Phys. {\bf A517}, 275 (1990).
\bibitem{block}
 V. G. Soloviev, Kgl. Dan. Vid. Selsk. Mat. Fys. Skr. {\bf 1}, 238 (1961).
\bibitem{t05}
J. Terasaki, J. Engel, M. Bender, J. Dobaczewski, W. Nazarewicz,
and M. Stoitsov, Phys. Rev. C {\bf 71}, 034310 (2005).
\bibitem{svbg13}
 A. P. Severyukhin, V. V. Voronov, I. N. Borzov and Nguyen Van Giai,
 Rom. Journ. Phys. {\bf 58}, 1048 (2013).
\bibitem{Suhonen}
 J. Suhonen, {\it From Nucleons to Nucleus} (Springer-Verlag, Berlin, 2007).
\bibitem{ebnds99}
J. Engel, M. Bender, J. Dobaczewski, W. Nazarewicz, and R. Surman,
Phys. Rev. C {\bf 60}, 014302 (1999).
\bibitem{ps72}
A. C. Pappas and T. Sverdrup, Nucl. Phys. {\bf A188}, 48 (1972).
\bibitem{sag11}
C.L. Bai, H.Q. Zhang, H. Sagawa, X.Z. Zhang, G. Col\`o and F.R.
Xu, Phys. Rev. {\bf C83}, 054316 (2011).
\bibitem{mb13}
F. Minato and C. L. Bai, Phys. Rev. Lett. {\bf 110}, 122501 (2013);
Phys. Rev. Lett. {\bf 116}, 089902(E) (2016).
\bibitem{ame2012}
M. Wang, G. Audi, A. H. Wapstra, F. G. Kondev, M. MacCormick, X.
Xu, and B. Pfeiffer, Chin. Phys. C {\bf 36}, 1603 (2012).
\bibitem{Jungclaus07}
A. Jungclaus, L. C\'{a}ceres, M. G\'{o}rska, M. Pf\"{u}tzner, S.
Pietri, E. Werner-Malento, H. Grawe, K. Langanke, G.
Mart\'{i}nez-Pinedo, F. Nowacki, A. Poves, J. J.
Cuenca-Garc\'{i}a, D. Rudolph, Z. Podolyak, P. H. Regan, P.
Detistov, S. Lalkovski, V. Modamio, J. Walker, P. Bednarczyk, P.
Doornenbal, H. Geissel, J. Gerl, J. Grebosz, I. Kojouharov, N.
Kurz, W. Prokopowicz, H. Schaffner, H. J.Wollersheim, K. Andgren,
J. Benlliure, G. Benzoni, A. M. Bruce, E. Casarejos, B. Cederwall,
F. C. L. Crespi, B. Hadinia, M. Hellstr\"{o}m, R. Hoischen, G.
Ilie, J. Jolie, A. Khaplanov, M. Kmiecik, R. Kumar, A. Maj, S.
Mandal, F. Montes, S. Myalski, G. S. Simpson, S. J. Steer, S.
Tashenov, and O. Wieland, Phys. Rev. Lett. {\bf 99}, 132501
(2007).
\bibitem{Ilieva14}
S. Ilieva, M. Th\"{u}rauf, Th. Kr\"{o}ll, R. Kr\"{u}cken, T.
Behrens, V. Bildstein, A. Blazhev, S. B\"{o}nig, P. A. Butler, J.
Cederk\"{a}ll, T. Davinson, P. Delahaye, J. Diriken, A.
Ekstr\"{o}m, F. Finke, L. M. Fraile, S. Franchoo, Ch. Fransen, G.
Georgiev, R. Gernh\"{a}user, D. Habs, H. Hess, A. M. Hurst, M.
Huyse, O. Ivanov, J. Iwanicki, P. Kent, O. Kester, U. K\"{o}ster,
R. Lutter, M. Mahgoub, D. Martin, P. Mayet, P. Maierbeck, T.
Morgan, O. Niedermeier, M. Pantea, P. Reiter, T. R. Rodr\'{i}guez,
Th. Rolke, H. Scheit, A. Scherillo, D. Schwalm, M. Seidlitz, T.
Sieber, G. S. Simpson, I. Stefanescu, S. Thiel, P. G. Thirolf, J.
Van de Walle, P. Van Duppen, D. Voulot, N. Warr, W. Weinzierl, D.
Weisshaar, F. Wenander, A. Wiens, and S. Winkler, Phys. Rev. C
{\bf 89}, 014313 (2014).
\bibitem{Kautzsch00}
T. Kautzsch, W. B. Walters, M. Hannawald, K.-L. Kratz, V. I.
Mishin, V. N. Fedoseyev, W. B\"{o}hmer, Y. Jading, P. Van Duppen,
B. Pfeiffer, A. W\"{o}hr, P. M\"{o}ller, I. Kl\"{o}ckl, V.
Sebastian, U. K\"{o}ster, M. Koizumi, J. Lettry, H. L. Ravn, and
the ISOLDE Collaboration, Eur. Phys. J. A {\bf 9}, 201 (2000).
\bibitem{Danchev11}
M. Danchev, G. Rainovski, N. Pietralla, A. Gargano, A. Covello, C.
Baktash, J. R. Beene, C. R. Bingham, A. Galindo-Uribarri, K. A.
Gladnishki, C. J. Gross, V. Yu. Ponomarev, D. C. Radford, L. L.
Riedinger, M. Scheck, A. E. Stuchbery, J. Wambach, C.-H. Yu, and
N. V. Zamfir, Phys. Rev. C {\bf 84}, 061306(R) (2011).
\bibitem{Varner05}
R. L. Varner, J. R. Beene, C. Baktash, A. Galindo-Uribarri, C. J.
Gross, J. Gomez del Campo, M. L. Halbert, P. A. Hausladen, Y.
Larochelle, J. F. Liang, J. Mas, P. E. Mueller, E. Padilla-Rodal,
D. C. Radford, D. Shapira, D. W. Stracener, J.-P. Urrego-Blanco,
and C.-H. Yu, Eur. Phys. J. A {\bf 25}, 391 (2005).
\bibitem{Watanabe13}
H. Watanabe, G. Lorusso, S. Nishimura, Z. Y. Xu, T. Sumikama,
P.-A. S\"{o}derstr\"{o}m, P. Doornenbal, F. Browne, G. Gey, H. S.
Jung, J. Taprogge, Zs. Vajta, J. Wu, A. Yagi, H. Baba, G. Benzoni,
K. Y. Chae, F. C. L. Crespi, N. Fukuda, R. Gernh\"{a}user, N.
Inabe, T. Isobe, A. Jungclaus, D. Kameda, G. D. Kim, Y. K. Kim, I.
Kojouharov, F. G. Kondev, T. Kubo, N. Kurz, Y. K. Kwon, G. J.
Lane, Z. Li, C.-B. Moon, A. Montaner-Piz\'{a}, K. Moschner, F.
Naqvi, M. Niikura, H. Nishibata, D. Nishimura, A. Odahara, R.
Orlandi, Z. Patel, Zs. Podoly\'{a}k, H. Sakurai, H. Schaffner, G.
S. Simpson, K. Steiger, H. Suzuki, H. Takeda, A. Wendt, and K.
Yoshinaga, Phys. Rev. Lett. {\bf 111}, 152501 (2013).
\bibitem{Mach95}
H. Mach, D. Jerrestam, B. Fogelberg, M. Hellstr\"{o}m, J. P.
Omtvedt, K. I. Erokhina, and V. I. Isakov, Phys. Rev. C {\bf 51},
500 (1995).
\bibitem{C07}
J. J. Cuenca-Garc\'{\i}a, G. Mart\'{\i}nez-Pinedo, K. Langanke, F. Nowacki,
 and I. N. Borzov,  F. Montes, Eur. Phys. J. A {\bf 34}, 99 (2007).
\bibitem{B14} I. N. Borzov, {\it Fission and Properties of Neutron-Rich Nuclei}, edited by  J. H. Hamilton,
A. V. Ramayya (World Scientific, Singapore, 2014), p. 530.
%
\end{thebibliography}
\end{document}